\documentclass[runningheads]{llncs}
\usepackage[T1]{fontenc}
\usepackage{graphicx,verbatim}
\usepackage{subfig}
\usepackage{xcolor}

\newcommand{\revise}[1]{\textcolor{black}{#1}}

\begin{document}
%

\title{Intelligent Virtual Sonographer (IVS): Enhancing Physician-Robot-Patient Communication}

%

\author{Tianyu Song\inst{1,2}$^{*}$ \and
Feng Li\inst{1,2}$^{*}$ \and
Yuan Bi\inst{1,2} \and
Angelos Karlas\inst{3} \and
Amir Yousefi\inst{4} \and
Daniela Branzan\inst{3} \and
Zhongliang Jiang\inst{1,2} \and
Ulrich Eck\inst{1} \and
Nassir Navab\inst{1,2}}

%
\authorrunning{T. Song and F. Li et al.}
\titlerunning{Intelligent Virtual Sonographer (IVS)}
%
\institute{Chair for Computer-Aided Medical Procedures and Augmented Reality, Technical University of Munich, Munich, Germany \and
Munich Center for Machine Learning, Munich, Germany \and
Department for Vascular and Endovascular Surgery, Rechts der Isar University Hospital, Technical University of Munich, Munich, Germany \and
Clinic for Vascular Surgery, Helios Klinikum München West, Munich, Germany \\
\email{zl.jiang@tum.de}\\
}
\begingroup\def\thefootnote{$*$}\footnotetext{These authors contributed equally to this work.}\endgroup


\maketitle              
\begin{abstract}
The advancement and maturity of large language models (LLMs) and robotics have unlocked vast potential for human-computer interaction, particularly in the field of robotic ultrasound. While existing research primarily focuses on either patient-robot or physician-robot interaction, the role of an intelligent virtual sonographer (IVS) bridging physician-robot-patient communication remains underexplored. This work introduces a conversational virtual agent in Extended Reality (XR) that facilitates real-time interaction between physicians, a robotic ultrasound system(RUS), and patients. The IVS agent communicates with physicians in a professional manner while offering empathetic explanations and reassurance to patients. Furthermore, it actively controls the RUS by executing physician commands and transparently relays these actions to the patient. By integrating LLM-powered dialogue with speech-to-text, text-to-speech, and robotic control, our system enhances the efficiency, clarity, and accessibility of robotic ultrasound acquisition. This work \revise{constitutes a first step toward understanding} how IVS can bridge communication gaps in physician-robot-patient interaction, providing more control and therefore trust into physician-robot interaction while improving patient experience and acceptance of robotic ultrasound. \revise{The code is available at \url{https://github.com/stytim/IVS}}


\end{abstract}

\section{Introduction}\label{sec:Introduction}

Advancements in robotic ultrasound systems (RUSs) ~\cite{jiang2023robotic,bi2024machine,jiang2024intelligent} and telemedicine~\cite{roth2021real,eck2023real} are transforming medical diagnostics by enabling automated and remote procedures. These technologies enhance accessibility to ultrasound imaging, particularly in rural or underserved areas, where trained sonographers may be unavailable. However, the integration of robotic ultrasound systems into clinical practice presents new challenges in communication, decision-making, and user acceptance~\cite{eilers2023importance}. A crucial aspect of this integration is the interaction between physicians, robotic systems, and patients —a communication dynamic that is often overlooked in existing research.

While prior studies have explored virtual agents for either patient engagement~\cite{bickmore2009taking,walker2020developing,song2025enhancing} or physician assistance~\cite{killeen2024intelligent,killeen2024take}, there remains a significant research gap in designing intelligent agents introduced in this work that facilitate communication among all three entities: the physician, the RUS, and the patient. This gap is critical, as effective coordination among these stakeholders is essential for diagnostic efficiency, physician trust, and patient acceptance. Without a structured communication framework, physicians may struggle to effectively convey instructions to the robotic system, patients may have difficulty understanding the physician’s intent, and physicians may misinterpret patient concerns. This lack of seamless multidirectional communication can lead to errors in robotic control, patient uncertainty, and decreased trust in the procedure.

Recent research in Embodied Conversational Agents (ECAs)~\cite{cassell2000embodied,cassell2001embodied} has shown that virtual avatars with human-like behaviors, facial expressions, and speech synchronization enhance trust, engagement, and communication effectiveness, particularly in healthcare applications. ECAs have been used to guide patients through medical procedures~\cite{bickmore2005establishing,ter2020design} and support physician decision-making~\cite{martinez2017embodied}, demonstrating their potential in clinical interactions. However, existing ECAs typically focus on single-user interactions, either with the patient or the physician, rather than facilitating real-time multidirectional communication between multiple stakeholders and a robotic system.

Building on this foundation, we introduce an Intelligent Virtual Sonographer (IVS) designed to bridge the interaction gap between the physician, the RUS, and the patient. This agent operates in an Extended Reality (XR) environment, offering assistance and procedural control for the physician, and providing empathetic guidance and reassurance for patients. The agent facilitates bidirectional information exchange, ensuring that physicians receive real-time updates on patient conditions while patients remain informed about the robotic ultrasound procedure. Additionally, the agent enhances transparency and user trust by explicitly communicating the robotic system’s actions to both parties.

Our approach leverages recent advancements in large language models (LLMs) to enable real-time, natural language interaction. The IVS facilitates multidirectional communication by translating physician instructions into robotic ultrasound commands and relaying system updates back to the physician. It allows patients to request pauses or pressure adjustments while keeping them informed about the procedure. Additionally, it ensures effective physician-patient communication by relaying patient concerns to the physician and conveying the physician’s intent to the patient. In this paper, we present the design, implementation, and evaluation of IVS. We assess its impact in bridging physician-robot-patient interactions and analyze its usability, communication effectiveness, perceived intelligence, and overall user satisfaction. Additionally, our system runs entirely locally to address medical privacy concerns, utilizing a relatively small LLM that operates efficiently without requiring a data center, yet still achieves strong performance. Our findings contribute to the broader field of LLM-driven AI agents in medical applications, particularly in telemedicine and robotic ultrasound-assisted diagnostics.
\section{Method}\label{sec:Method}

\begin{figure}
    \centering
    \includegraphics[width=0.78\linewidth]{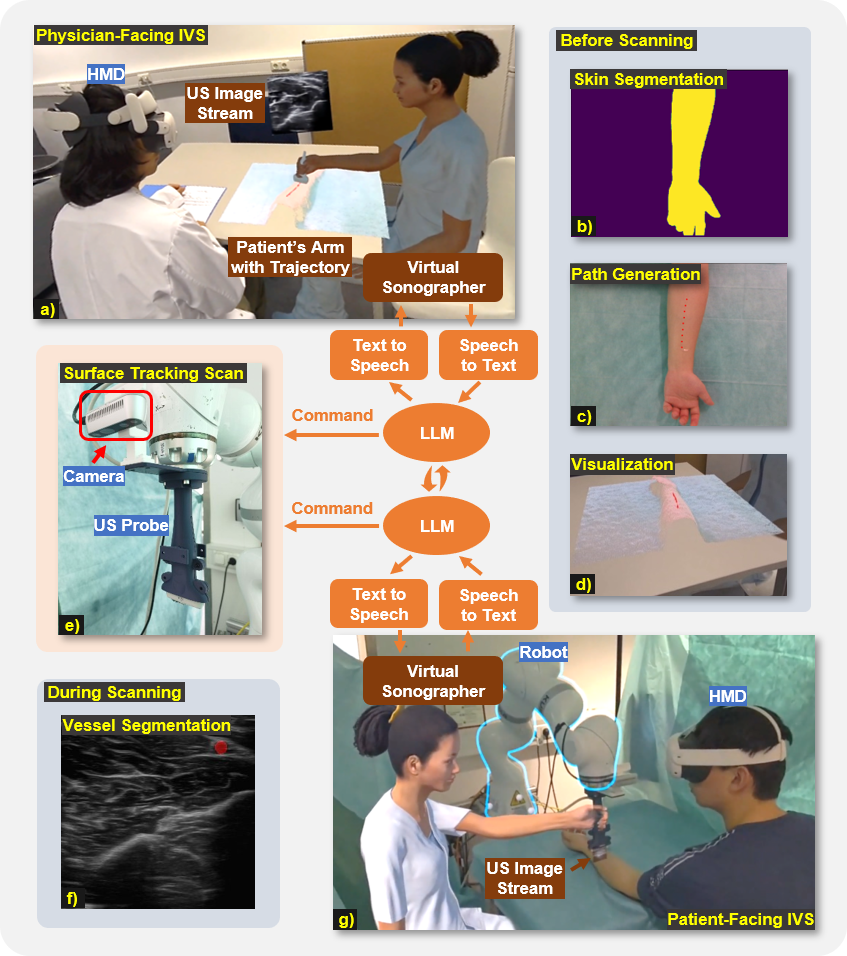}
    \caption{System overview. (a) IVS interacting with the physician in XR. (b)-(d) Automatic trajectory generation visualized in XR. (e) RUS hardware setup with tracking camera, ultrasound probe, and robot. (f) Real-time vessel segmentation on the ultrasound image. (g) IVS interacting with the patient. The straight arrow represents the data transmission flow, the curvy arrow shows the interaction.}
    \label{fig:pipeline}
\end{figure}

Our system implements an XR-driven robotic ultrasound workflow, as illustrated in Fig.~\ref{fig:pipeline}, allowing both the physician and patient to interact with virtual sonographers. The RUS autonomously plans and performs ultrasound scanning by extracting intent from natural conversations. The following sections provide a detailed overview of the RUS architecture and the IVS.

\subsection{Robotic Ultrasound System}
To enable automatic scanning, the robotic ultrasound system is designed with a KUKA LBR iiwa 14 R820 robotic arm and a Siemens ACUSON Juniper ultrasound machine. The Siemens 12L3 ultrasound probe is rigidly attached to the robot’s end-effector using a 3D-printed probe holder. Additionally, an Intel Realsense D435i depth camera is mounted between the probe and the robot end-effector (Fig.~\ref{fig:pipeline}(e)) to capture the 3D point cloud from the patient side, which is then reprojected for the physician’s visualization. The hand-eye calibration \cite{horaud1995hand} is used to transform the camera pose into the robot end-effector pose.

Before scanning, the physician can instruct the IVS system to display the patient's arm and the planned scanning trajectory. During this process, one RGB image along with ten consecutive depth images are acquired. To generate the scanning path, we employ the MediaPipe \cite{lugaresi2019mediapipe} framework to extract specific body features, including skin segmentation and hand landmark detection. First, the arm mask is obtained using skin segmentation \revise{with a trained model, SelfieMulticlass} (Fig.~\ref{fig:pipeline}(b)). To better determine the scanning direction, \revise{HandLandmarker model} is applied to identify the wrist point pose. Additionally, the arm mask’s skeleton is computed, starting from the wrist and extending to the widest part of the forearm. While these steps rely on the RGB image, generating the full scanning path requires 3D point information, which is extracted from the depth images. We capture ten depth images, extract the depth values at each selected path point, and apply a median filter to refine the depth estimation. Once the origins of each scanning point $ (t_x,\ t_y,\ t_z) $ are determined, 16 neighboring points for each selected point along the path are acquired and a plane is fitted to calculate the normal vector. This step ensures that the robot's end effector maintains the smallest possible angle with respect to the normal direction of each path point, allowing for smooth movement and precise surface tracking \cite{jiang2020automatic}. As a result, high-quality ultrasound images can be obtained. Finally, we compute the full 6D pose $ (t_x,\ t_y,\ t_z,\ r_x,\ r_y,\ r_z)$ in the robot base coordinate system for each path point (Fig.~\ref{fig:pipeline}(c)) and visualize the trajectory for the physician (Fig.~\ref{fig:pipeline}(d)).

During scanning, the RUS employs impedance control to optimize the contact between the ultrasound probe and the skin. The system can dynamically adjust its behavior—pausing, increasing pressure, or reducing pressure—to accommodate the physician’s specific needs. Additionally, the UNet model \cite{ronneberger2015u} \revise{trained on 3000 US images (Dice=0.954 $\pm$ 0.012, precision=0.942 $\pm$ 0.021)} is used to segment the blood vessels in real-time (Fig.~\ref{fig:pipeline}(f)), and the blood vessels' position is relayed to the physician-facing IVS to provide prompts for the physician. After completing the scan, the RUS signals the IVS to conclude the process.


\subsection{Intelligent Virtual Sonographer}

\subsubsection{Multidirectional Communication Framework.}

\begin{figure}
    \centering
    \includegraphics[width=0.8\linewidth]{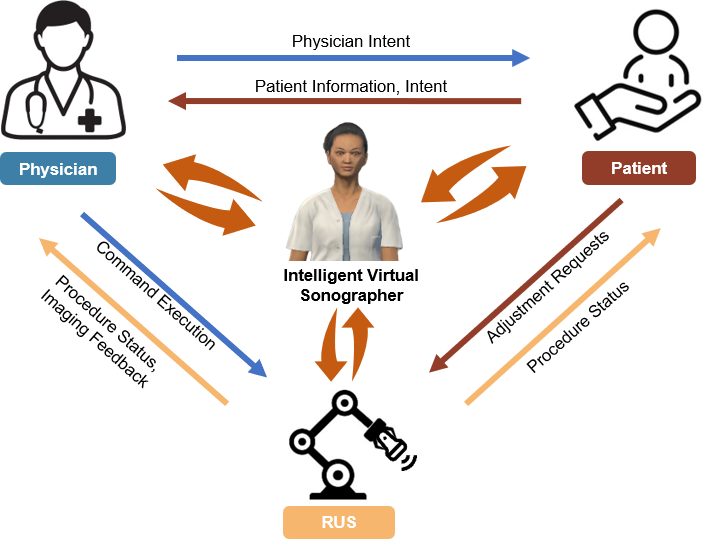}
    \caption{The IVS facilitates seamless interaction between the physician, robotic ultrasound system, and patient.}
    \label{fig:communication}
\end{figure}

Effective communication between the physician, RUS, and patient is essential for ensuring accurate diagnostics, efficient workflow, and patient comfort. The IVS facilitates this interaction through three key communication pathways, shown in Fig.~\ref{fig:communication}:

\textbf{Physician $\leftrightarrow$ Robot}: The physician issues verbal commands to control the RUS, ensuring accurate execution of scanning maneuvers. The robotic system provides real-time feedback on probe positioning and detected anatomical structures, such as vessel locations, helping the physician make informed decisions.

\textbf{Physician $\leftrightarrow$ Patient}: The IVS enhances physician-patient communication by conveying the physician’s intent and procedural instructions in an accessible manner. Simultaneously, it ensures the physician receives patient-specific information, including medical history, current condition, and patient preferences, supporting personalized care.

\textbf{Patient $\leftrightarrow$ Robot}: Patients can interact with the RUS by expressing discomfort or requesting adjustments, such as reducing probe pressure or pausing the scan. The robotic system, in turn, provides real-time updates on its status, reassuring the patient about the ongoing procedure.

To enable multidirectional communication, the IVS must support simultaneous interactions between the physician, patient, and RUS. Physician-patient communication is inherently asynchronous—the physician and patient may request or provide information at different times, requiring the IVS to handle simultaneous interactions without blocking one conversation for the other. Running two independent LLM instances allows the IVS to manage both interactions in parallel, ensuring that the patient-facing conversation does not delay or interfere with the physician’s workflow and vice versa. To achieve this, the IVS utilizes two separate instances of a LLM, each running Meta Llama 3.1 8B model~\cite{dubey2024llama} without fine-tuning. These LLM instances were deployed locally on an NVIDIA IGX Orin system equipped with an RTX 6000 Ada GPU.
To facilitate dual-role interaction, each LLM instance was prompted differently: one was designated for professional communication with the physician, while the other engaged in empathetic conversation with the patient, ensuring clear and supportive explanations.
When the physician requested patient-specific information, the physician-facing LLM relayed the query to the patient-facing LLM if it lacked the required data. The patient-facing LLM then asked the patient for a response, after which it relayed the information back to the physician-facing LLM and notified the physician accordingly. Beyond conversation, the IVS also interpreted verbal physician instructions to execute robotic ultrasound actions. Using an approach similar to Xu \emph{et al.}~\cite{xu2024transforming}, the physician-facing LLM converted spoken commands into robotic control API calls, enabling the IVS to execute the scanning, adjust ultrasound probe pressure, pause or resume the procedure. The IVS also transparently communicated these actions to the patient, reinforcing trust and clarity in robotic-assisted diagnostics.

\subsubsection{XR Visualization and Embodied Interaction.}
The visual representation of the IVS was developed in Unity 2022.3.55, utilizing avatars from the Microsoft Rocketbox library~\cite{gonzalez2020rocketbox}. Predefined animations, such as sitting with breathing and blinking, were baked into the system, while dynamic animations—including head turning during conversations and procedural grasping of the ultrasound probe—were generated in real time using Final IK~\cite{finalik}. Facial lip-syncing animations were driven by SALSA~\cite{salsa}, enabling mouth movements synchronized with speech output. The Unity application was executed on a Windows PC equipped with an NVIDIA RTX 4060 GPU and visualized on a Meta Quest 3 headset using Quest Link mode.
For speech processing, the system employed the OpenAI Whisper base model for speech-to-text (STT), running on the same PC to transcribe user speech for the LLM. The LLM-generated responses were then synthesized into speech using Kokoro~\cite{kokoro}, an open-source text-to-speech (TTS) model, which was deployed on the same server as the LLM. This pipeline enabled real-time spoken communication between the IVS and users.
\section{Experiment and Results}\label{sec:Experiment}

\subsection{User Study}
To evaluate the IVS, we conducted a user study assessing its communication accuracy, perceived intelligence, interaction quality, usability, and overall satisfaction. The study aimed to measure both objective communication effectiveness and subjective user experience through structured participant feedback.

While prior work by Song \emph{et al.}~\cite{song2025enhancing} has explored the effects of a virtual agent designed for patient interaction, our study primarily focuses on the physician-facing IVS and its role in facilitating physician-robot-patient communication. To this end, we recruited 14 participants (4 female, 10 male), ranging in age from 24 to 42 years (mean age: 31.6 years). Among them, 7 were medical doctors with an average of 6.7 years of professional experience\revise{, while the remaining were biomedical engineers}. The study was conducted in accordance with the Declaration of Helsinki. Ethical approval was obtained from the ethics committee of the \revise{Technical University of Munich}. To ensure a controlled and consistent evaluation, participants took on the role of physicians, while the authors of this paper acted as patients, adopting pre-defined patient personas with different names, ages, and medical histories for each physician participant. 

In the user study, the physician was required to retrieve the patient-specific information, including name, age, and medical history, which the IVS relayed during the interaction. To assess communication accuracy, the information provided by the patient was compared to what the physician received through the IVS. Beyond information retrieval, the physician needed to initiate the scanning process verbally and interact with the IVS to control the RUS. Through natural conversation, the physician could instruct the IVS to adjust probe pressure, pause the movement to examine the ultrasound image more carefully, and resume scanning as needed. Additionally, the patient could also interact with the IVS by requesting adjustments to the probe pressure or pausing the procedure, simulating situations where they might feel discomfort during the examination. These tasks evaluated whether the IVS effectively facilitated multidirectional intent communication, ensuring that the physician’s instructions were accurately executed by the robot, that the patient’s concerns were properly relayed and addressed, and that both the physician and patient correctly understood each other’s intent during the interaction.



\subsection{Quantitative Evaluation}

For the IVS to support natural and fluid conversation, system responsiveness is critical.
In our setup, the IVS achieved an average frame rate of 72 FPS at 2,064 $\times$ 2,208 pixels per eye. On average, the latency for STT was 80 ms, while LLM response generation was 860 ms. The TTS synthesis introduced an additional 150 ms delay, resulting in a total conversational latency of 1.09 seconds per turn. 


In terms of communication effectiveness, the IVS achieved an accuracy of 90.48\% in relaying patient-specific information to the physician. Additionally, the IVS accurately executed 85.71\% of patient-requested actions, such as adjusting probe pressure or pausing the procedure. For physician-requested actions, including scanning maneuvers and robotic adjustments, the system achieved an accuracy of 92.86\%. Despite these high accuracy rates, some failure cases were observed during the user study. Occasionally, the LLM hallucinated patient-specific information, generating responses from example patients in the prompt instead of accurately retrieving the correct participant information.

\subsection{Subjective Ratings}
\begin{figure}
    \centering
    \includegraphics[width=\linewidth]{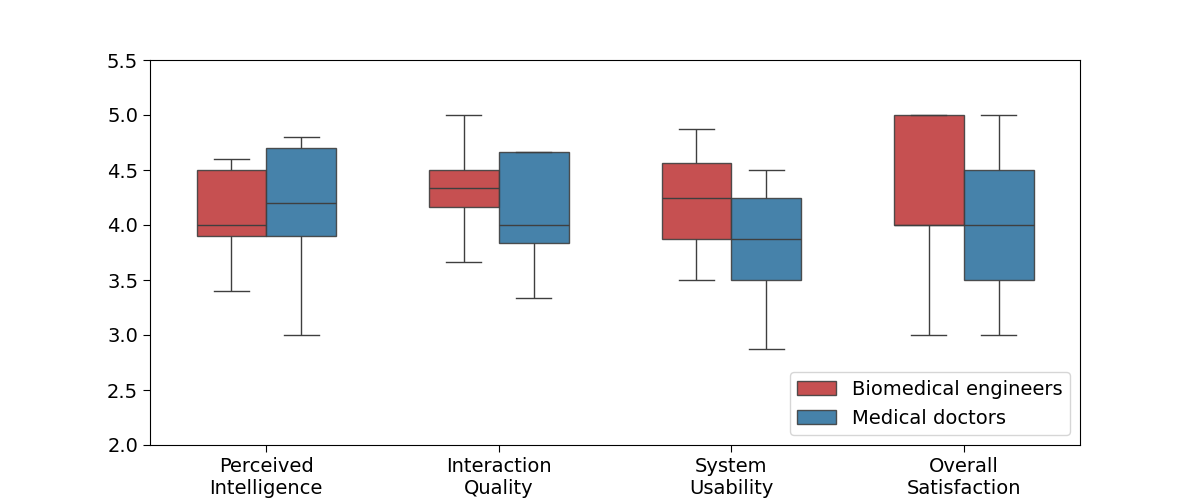}
    \caption{Subjective ratings of \revise{biomedical engineers} and \revise{medical doctors}.}
    \label{fig: user study}
\end{figure}

The subjective ratings are categorized into four aspects: perceived intelligence, interaction quality, system usability, and overall satisfaction, in a 5-point Likert scale\revise{, where higher scores indicate better performance}. 
The results, as shown in Fig.~\ref{fig: user study}, are presented separately for novice and physician participants (mean ± standard deviation).
Perceived intelligence was rated 4.11 $\pm$ 0.45 \revise{(median: 4)}, and 4.17 $\pm$ 0.64 \revise{(median: 4.2)} by physicians. 
Interaction quality received 4.33 $\pm$ 0.43 \revise{(median: 4.33)} from novices, and 4.14 $\pm$ 0.54 \revise{(median: 4)} from doctors.
System usability was rated 4.21 $\pm$ 0.51 \revise{(median: 4.25)} by novices but lower by physicians at 3.82 $\pm$ 0.59 \revise{(median: 3.86)}.
Overall satisfaction was 4.28 $\pm$ 0.75 \revise{(median: 4)} for novices, and 4.00 $\pm$ 0.81 \revise{(median: 4)} for physicians.
While both groups provided high ratings, the lower interaction, usability and satisfaction score from physicians suggested potential areas for further improvement.


\section{Discussion and Conclusion}\label{sec:Conclusion}


In this work, we introduced IVS, a dual-LLM-driven embodied conversational agent, designed to enhance physician-robot-patient communication by facilitating natural and interactive dialogue in robotic ultrasound. Even with a moderate-sized LLM without fine-tuning, IVS \revise{provides initial evidence of facilitating} communication, improving the efficiency and transparency of robotic ultrasound. However, challenges remain in latency and LLM reliability. While the system latency exceeds typical human response times~\cite{stivers2009universals}, it remains within an acceptable range for interactive medical applications~\cite{huq2024dialogue}. 
Another critical aspect is LLM hallucination, which risks generating incorrect clinical information. Our dual-instance LLM architecture mitigates this to some extent, but integrating medical knowledge bases and physician validation mechanisms could further enhance reliability. Despite these challenges, IVS received positive ratings, demonstrating its potential as a scalable, locally deployed AI agent in medical robotics. \revise{While current LLMs have limitations in complex medical contexts, rapid advancements in contextual understanding and reliability are underway, and our framework is positioned to benefit from these improvements. As IVS intelligence evolves, the interaction paradigm could also shift from low-level robot control to higher-level clinical communication, allowing physicians to request anatomical views rather than specifying probe manipulations—aligning with existing clinical workflows. Additionally, future integration of sensing modalities such as eye-tracking, heart rate, facial expression, or voice analysis may enable IVS to infer patient emotional states and relay them to physicians, enriching remote diagnostic interactions.} 
This novel approach paves the way for future advancements, encouraging the community to further develop AI-driven virtual agents that enhance trust, usability, and real-world impact in robotic-assisted diagnostics.





\begin{credits}
\subsubsection{\ackname} The authors wish to thank NVIDIA for their in-kind hardware and advisory support. This work was partly supported by the state of Bavaria through Bayerische Forschungsstiftung (BFS) under Grant AZ-1592-23-ForNeRo.

\subsubsection{\discintname}
The authors have no competing interests to declare that are
relevant to the content of this article. 
\end{credits}

%
%
%
\bibliographystyle{splncs04}
\bibliography{09_Bibliography}

\end{document}